\documentstyle[mnextra]{mn}
\seceqnum           
\input{epsf}
\def\etal{{\rm et al.} }
\def\kms  {\,{\rm km \, s^{-1}}}
\def\Mpc  {\,{\it h}^{-1}\, {\rm Mpc}}

\def\Msol {\,{\it h}^{-1}\, {\rm M_\odot}}
\def\hMsol {\,{\it h}^{-2}\, {\rm M_\odot}}
\def\Lsol {\,{\it h}^{-2}\, {\rm L_\odot}}

\def\d    {{\rm d}}
\def\i    {{\rm i}}
\def\j    {{\rm j}}
\def\k    {{\rm k}}
\def\g    {{\rm g}}
\def\p    {{\rm p}}
\def\q    {{\rm q}}
\def\r    {{\rm r}}
\def\ie {{\rm i.e. }}
\def\eg {{\rm e.g. }}
\def\log {{\rm log}}

\def\lsim{\mathrel{\hbox{\rlap{\hbox{\lower4pt\hbox{$\sim$}}}\hbox{$<$}}}}
\def\gsim{\mathrel{\hbox{\rlap{\hbox{\lower4pt\hbox{$\sim$}}}\hbox{$>$}}}}

\def\bj{b_{\rm J}}
\def\rf{r_{\rm F}}

\def\m@th{\mathsurround=0pt }
\def\eqalign#1{\null\,\vcenter{\openup1\jot \m@th
 \ialign{\strut\hfil$\displaystyle{##}$&$\displaystyle{{}##}$\hfil
 \crcr#1\crcr}}\,}

\begin{document}
\title[Where are the stars?]
{Where are the stars?}
\author[V.R. Eke et al.]
{\parbox[t]\textwidth{
V.R. Eke$^{1}$,
C.M. Baugh$^{1}$,
Shaun Cole$^1$,
Carlos S. Frenk$^1$,
H.M. King$^{1}$,
John A. Peacock$^{2}$}
\vspace*{6pt} \\
{$^1$Department of Physics, University of Durham, South Road,
    Durham DH1 3LE, UK} \\
{$^2$Institute for Astronomy, University of Edinburgh, Royal 
       Observatory, Blackford Hill, Edinburgh EH9 3HJ, UK}\\
}
\maketitle
\begin{abstract}
The Two degree Field Galaxy Redshift Survey (2dFGRS) is used in
conjunction with the Two Micron All Sky Survey (2MASS) Extended Source
Catalogue (XSC) to study the near-infrared light and stellar 
mass content of the local Universe. Mock galaxy catalogues,
constructed from cosmological N-body simulations and semi-analytical galaxy
formation models, are used to gauge the accuracy with which quantities
can be recovered. The mean luminosity densities of the Universe
are found to be $\bar\rho_J=(3.57\pm0.11)\times10^8 h
L_\odot{\rm Mpc}^{-3}$ and $\bar\rho_{K_S}=(7.04\pm0.23)\times10^8 h
L_\odot{\rm Mpc}^{-3}$ (statistical uncertainty only, and not
accounting for the 2MASS low surface brightness incompleteness). 
Using the 2dFGRS Percolation-Inferred Galaxy Group (2PIGG) catalogue, the 
group mass-to-light ratio in the $K_S$ band is found to increase by a
factor of $\sim 3$ when going from groups with total $\bj$-band
luminosities of $3 \times 10^{10} \Lsol$ to the richest
clusters. These clusters have typical dynamical mass-to-light ratios of
$\Upsilon_K \approx 80 h \Upsilon_\odot$. Galaxy luminosities 
are used to estimate stellar masses.
Taking into account the bias introduced by uncertainties in estimating
galaxy stellar masses, a value of 
$\Omega_{\rm stars}h=(0.99\pm0.03)\times10^{-3}$ is measured, assuming 
that a Kennicutt stellar initial mass function (IMF) is applicable to all
galaxies. Changing this to a Salpeter stellar IMF gives 
$\Omega_{\rm stars}h \approx 2.1\times10^{-3}$. The 2PIGGs are then
used to study the distribution of the stellar content of the local Universe
among groups of different size. The three main conclusions are:\\
(1) a slowly rising stellar mass-to-$K_S$
band light ratio is found with the clusters having the largest value
of $\sim 0.6\Upsilon_\odot$,\\ 
(2) in contrast, the fraction of mass in
stars decreases with increasing group size, reaching $\sim
5\times10^{-3}h$ for the rich clusters, and \\
(3) in answer to the question posed in the title, most stellar mass is
contained in Local Group-sized objects ($M\sim2\times10^{12}\Msol$)
with only $\sim 2$ per cent in clusters with $M\gsim5\times10^{14}\Msol$.
\end{abstract}

\section{Introduction}

Understanding the impact that environment has on a galaxy's ability to
form stars is one of the main goals of cosmology. A very
pertinent observational constraint is the manner in which stellar
mass is distributed between galaxy systems of different size. This
paper describes a measurement of this constraint, using a combination
of the Two Micron All Sky Survey (2MASS) Extended Source Catalogue 
(Jarrett \etal 2000) and the 2dFGRS (Colless \etal 2001, 2003)
Percolation-Inferred Galaxy Group (2PIGG) catalogue (Eke \etal 2004).

By-products of this quest, which are also reported here, include a
recalculation of the local galaxy near-IR luminosity and stellar mass
functions with the largest sample of galaxies yet used for this
purpose. Mock galaxy catalogues are used in order to help quantify
systematic uncertainties and biases. In addition,
the dependence on group size of the stellar mass fraction and of the group
mass-to-light and stellar mass-to-light ratios are studied and
compared with previous results.

Section~\ref{sec:data} contains a description of the data used
in this analysis. The near-IR luminosity functions and mean luminosity
densities are calculated in Section~\ref{sec:irlf} and compared with
recent determinations by Cole \etal (2001; hereafter C01), 
Kochanek \etal (2001)
and Bell \etal (2003). Group total mass-to-light ratios in the $K_S$
band are reported in Section~\ref{sec:molk}. These results extend the
range of group sizes over which near-IR mass-to-light ratios have been
measured, relative to the studies of Kochanek \etal (2003), Lin, Mohr
\& Stanford (2003; hereafter LMS03) and Rines \etal (2004).
Section~\ref{sec:starsest} contains a
description of the procedure used to calculate stellar masses given a
set of galaxy luminosities in different wavebands. The accuracy of
these measurements is gauged using mock galaxy catalogues produced
from a combination of a dark matter N-body simulation and a semi-analytical
galaxy formation model. This information is then used in 
Section~\ref{sec:starst}, where the total galaxy stellar mass function
and $\Omega_{\rm stars}$ are estimated.

The stellar content of the galaxy population is split by group size in
Section~\ref{sec:starsg}, where group stellar mass-to-light ratios and
stellar mass fractions are presented and compared with those found by LMS03.
Finally, Section~\ref{sec:where} provides the constraint
alluded to at the start of the Introduction. Throughout this paper, it
is assumed that $\Omega_{\rm m}=0.3$, $\Omega_\Lambda=0.7$ and the Hubble
constant is written as $H_{0}=100h\kms{\rm Mpc^{-1}}$.

\section{Observational data}\label{sec:data}

\subsection{2dFGRS}\label{ssec:2df}

The two contiguous patches of sky covered by the 2dFGRS contain
approximately $190\,000$ galaxies with a median redshift of $0.11$. For
the purposes of the work performed in this paper, only the $\sim
109\,000$ galaxies at $z<0.12$ were used. At higher redshifts, less 
than half of the $\bj$-band luminosity of typical groups
is contained within galaxies above the survey flux limit. The parent
catalogue for the 2dFGRS has a very well quantified completeness.
Norberg \etal (2002) and
Cross \etal (2004) estimate that only $\sim 9$ per cent of the $\bj$-band
luminosity is missing, while Driver \etal (2004) conclude from
the MGC that
the 2dFGRS-inferred luminosity density is $\sim 3$ per cent low due to
missing low surface brightness galaxies (at all luminosities).

In Sections~\ref{sec:molk},~\ref{sec:starsg} and~\ref{sec:where},
where the galaxies 
are split by host group size, the 2PIGG catalogue (Eke \etal 2004) has
been used to define the galaxy groups and their properties. Note that in
Section~\ref{sec:where}, where the distribution of stellar mass
throughout the group population is studied, single galaxies are
also defined as groups. While such groups do not have a dynamically
estimated mass, they still have a well defined total group
luminosity.

\subsection{2MASS}\label{ssec:2mass}

The near-IR data were obtained from the 2MASS All-Sky Extended Source 
Catalogue (Jarrett \etal 2000), using the database available at
http://irsa.ipac.caltech.edu/cgi-bin/Gator/nph-dd.
A total of 43\,553
$z<0.28$ 2MASS galaxies were matched with those in the two contiguous
patches of the 2dFGRS. The median redshift of the 2dFGRS galaxies with
2MASS matches is $\sim 0.095$, slightly lower than the median of all
2dFGRS galaxies ($\sim 0.11$).
\begin{figure}
\centering
\centerline{\epsfxsize=8.6cm \epsfbox{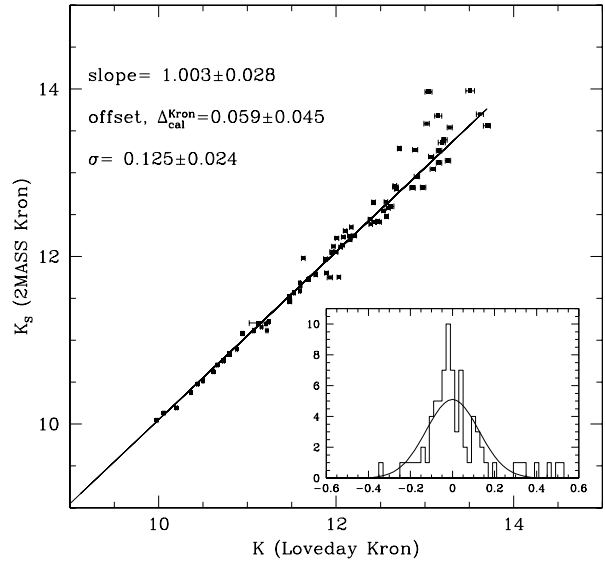}}
\caption{Comparison of the 2MASS $K$-band Kron magnitudes with those
measured by Loveday (2000). A straight line traces the least-squares
fit to the data. The inset histogram shows the
distribution of magnitude differences between the 2MASS and Loveday
measurements, which is interpreted as the distribution of errors in
the 2MASS values. The best-fitting Gaussian is also shown, and its
parameters are given in the legend.}
\label{fig:2mass1}
\end{figure}

The total galaxy $J$-band magnitude was estimated from the $J$-band
Kron measurement (corrected for Galactic extinction), 
including an offset of $-0.1$mag to account for flux
outside the Kron aperture. This is approximate because of the variety
of different surface brightness distributions among the galaxies,
but it is a reasonable typical correction (C01). The total galaxy
$K_S$-band value is then inferred using the total $J$-band magnitude
and the $J-K_S$ colour within the 20 mag per square arcsec isophotal fiducial
elliptical aperture. Following C01, the $K_S$-band Kron
magnitudes (without the $-0.1$mag shift) were compared with the
deeper pointed observations of Loveday (2000). Fig.~\ref{fig:2mass1}
shows that the 2MASS $K_S$-band Kron values inferred from the $J$-band
Kron magnitude and the isophotal colour are essentially unbiased, and
have a scatter about the Loveday measurements that depends strongly on
magnitude. This represents an improvement over the first 2MASS data release
used by C01, who needed to apply an offset to match the Loveday data.
For the whole sample of 80 galaxies, the distribution of
errors, shown in the inset histogram, has a best-fitting Gaussian
standard deviation of $0.125$mag.

\begin{figure}
\centering
\centerline{\epsfxsize=8.6cm \epsfbox{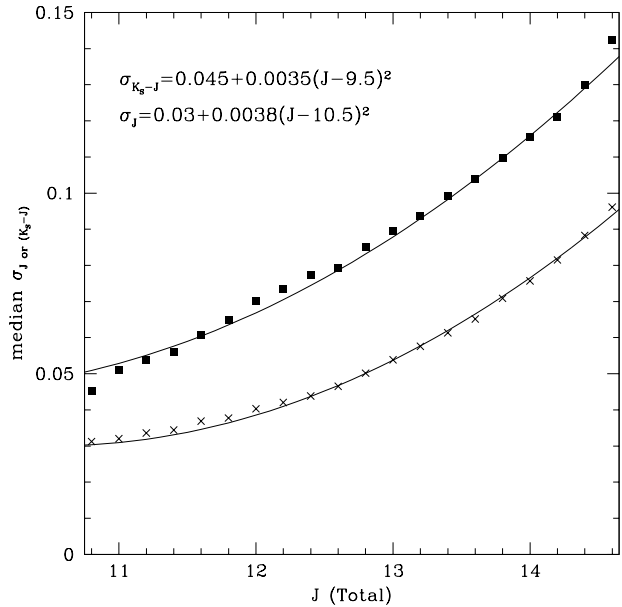}}
\caption{The errors on the 2MASS photometry are shown as functions of
galaxy $J$-band total magnitude. Crosses and filled squares represent
the medians of the $J$-band and $K_S-J$ $1\,\sigma$ uncertainties in
each bin, \ie the typical errors. The
individual galaxy errors used to construct these medians are those contained in
the 2MASS database. The two lines trace the quadratic fits given in
the panel.}
\label{fig:2masserrs}
\end{figure}
When creating mock catalogues with near-IR galaxy magnitudes, it is
necessary to include the observational errors in order to determine the
systematic bias that they introduce. The medians of the binned
errors on the $J$-band total magnitude and the $J-K_S$ colour are shown in
Fig.~\ref{fig:2masserrs} as functions of the total $J$-band
magnitude. These have been constructed using the $1\,\sigma$
uncertainties returned from the 2MASS database. As was apparent in
Fig.~\ref{fig:2mass1}, the observational errors are greater for the
fainter galaxies.

The $k+e$ corrections adopted for the $J$ and $K_S$ bands are
\begin{equation}
k+e=-\frac{(z+4z^2)}{1+9z^{2.5}}
\end{equation}
and
\begin{equation}
k+e=-\frac{(z+15z^2)}{1+9z^{2.5}}
\end{equation}
respectively. 
These are fits to the mean of the $k+e$ corrections computed for
individual galaxies using stellar population models. A detailed
description of how the $k+e$ corrections are calculated can be found in
C01. In brief, look-up tables of $b_{J}$, $\rf$, $J$
and $K$ magnitudes as a function of redshift were generated for a grid of
star formation histories, using the stellar population spectral energy
distributions provided by Bruzual \& Charlot (2003).
Observed galaxies were matched with theoretical star formation histories,
thus allowing a $k+e$ correction to be assigned to each galaxy using the
theoretical model. These simple models are used in
Section~\ref{sec:starsest} to assign stellar masses to galaxies.

\begin{figure}
\centering
\centerline{\epsfxsize=9cm \epsfbox{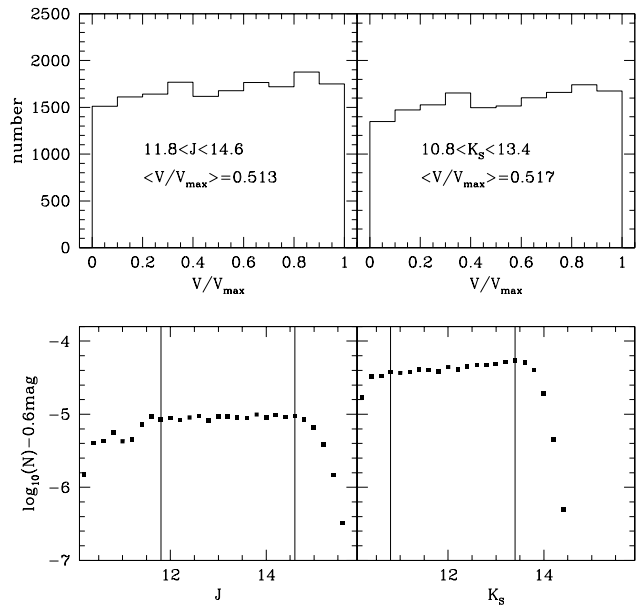}}
\caption{The top row shows the distributions of $V/V_{\rm max}$ for
the $J$- (left) and $K_S$-band (right panel) samples. The lower row
contains plots of the total number counts ($N$) as a function of magnitude for
these same two bands. Vertical lines illustrate the flux limits
applied to the survey in this paper.}
\label{fig:2mcomp}
\end{figure}

Note that the mean $V/V_{\rm max}=0.513$ for the 16\,922 galaxies
with $11.8<J<14.6$ and $z<0.12$. In this expression, $V$ represents
the volume surveyed out to the redshift of each particular galaxy, and
$V_{\rm max}$ is the maximum volume in which each galaxy could have
been detected. The $K_S$-band sample of 15\,664
galaxies with $10.8<K_S<13.4$ and $z<0.12$ has a mean 
$V(z_\i)/V(z_{\rm max,i})=0.517$. These values suggest that the
galaxy samples are not significantly incomplete (which would lead to
values lower than $0.5$) when these flux limits are applied. The top
panels in Fig.~\ref{fig:2mcomp} show the full distributions of
$V(z_\i)/V(z_{\rm max,i})$ for the samples used in the two bands, and
the lower panels show the differential number counts with the
Euclidean slope removed for all of the matched galaxies. These suggest
that there is no strong systematic depletion of the galaxies near
the adopted flux limits. Andreon (2002) and
Bell \etal (2003) find a small incompleteness in the 2MASS survey arising
from a deficit of faint, low surface brightness galaxies. While this
makes a significant difference to the luminosity and stellar mass
functions for the smaller galaxies, it has only a $\sim 20$ per cent 
impact upon
the total luminosity or stellar mass densities inferred from the
data. This will be discussed further in Section~\ref{sec:starst}.

\section{Luminosity functions}\label{sec:irlf}

\begin{figure}
\centering
\centerline{\epsfxsize=8.6cm \epsfbox{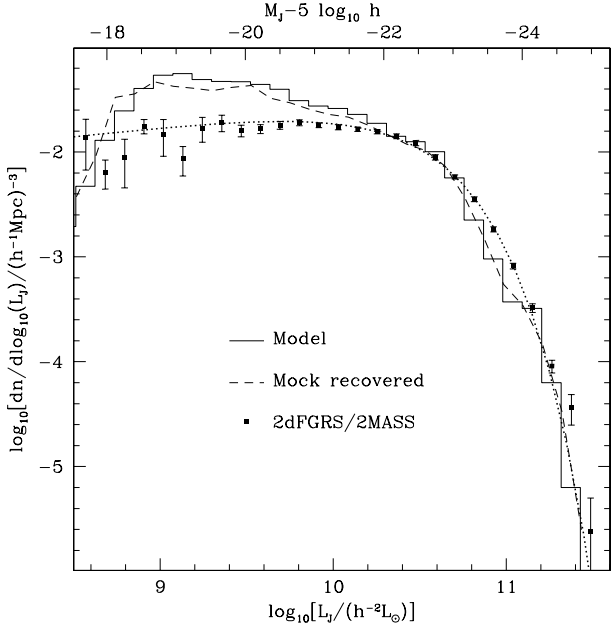}}
\caption{$J$-band luminosity functions for mock catalogues and
2dFGRS/2MASS data. The solid histogram traces the model galaxy $J$-band
luminosity function, whereas the dashed line shows the luminosity
function actually recovered from the
mock catalogue using a SWML estimator, normalised to agree with the
$1/V_{\rm max}$ luminosity function for
$10.1<\log_{10}[L_J/(\Lsol)]<11.2$. Results for the SWML estimate
from the real data are shown with the points, and the error bars are
inferred using a jack-knife method. The
dotted line traces the best-fitting Schechter function
derived from a $\chi^2$ fit to the SWML estimate over the same range
of luminosities used above. The
Schechter function parameters are given in Table~\ref{tab:schecl}.}
\label{fig:schecj}
\end{figure}

\begin{figure}
\centering
\centerline{\epsfxsize=8.6cm \epsfbox{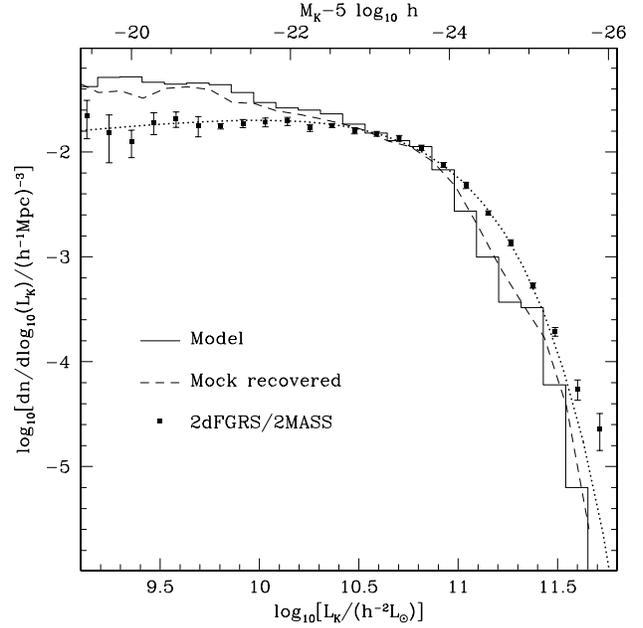}}
\caption{The equivalent of Fig.~\ref{fig:schecj} for the $K_S$-band
luminosity functions. The SWML estimator has been normalised
by matching to the $1/V_{\rm max}$ luminosity function at
$10.4<\log_{10}[L_{K_S}/(\Lsol)]<11.4$.}
\label{fig:scheck}
\end{figure}

\begin{table*}
\begin{center}
\caption{Best-fitting Schechter function parameters for the different
galaxy luminosity functions. The normalisations come from a $\chi^2$
fit to the $1/V_{\rm max}$ estimate of the abundances of the most
luminous galaxies. 
Also listed are the adopted solar absolute magnitudes in the two bands and the
mean Universal luminosity densities implied
by the SWML functions. Only statistical uncertainties are quoted.
The larger systematic uncertainties are discussed in the text. No
account has been taken of the systematic impact of the errors in the
2MASS magnitudes. However, this effect is smaller than the quoted
statistical uncertainties.}
\begin{tabular}{lllllll} \hline
band & ~~~~~~~~$M_*$ &
$~~~~L_*/(h^{-2}L_\odot)$ & ~~~~~~~~$\alpha$ & ~~~$\phi_*/(h^{-1}{\rm Mpc})^{-3}$ & $M_\odot$ & $\bar{\rho}/(h L_\odot{\rm Mpc}^{-3})$ \\
\hline
$J$&$-22.39\pm0.05$&$(2.81\pm0.12)\times10^{10}$&$-0.82\pm0.06$&$(1.39\pm0.06)\times10^{-2}$&3.73&$(3.57\pm0.11)\times10^8$\\
$K_S$&$-23.43\pm0.04$&$(5.36\pm0.22)\times10^{10}$&$-0.81\pm0.07$&$(1.43\pm0.08)\times10^{-2}$&3.39&$(7.04\pm0.23)\times10^8$\\
\hline
\end{tabular}
\label{tab:schecl}
\end{center}
\end{table*}
\begin{table}
\caption{The near-IR luminosity functions recovered with the real
data. Luminosities are measured in $\Lsol$ and the quoted abundances are
$\d n/\d\log_{10}L$ in $(h^{-1}{\rm Mpc})^{-3}$.}
\begin{tabular}{lll} \hline
$\log_{10}(L)$ & ~~~~~~$J$ band & ~~~~~~~$K_S$ band \\
\hline
$\hspace{0.15cm} 8.50$&$(1.08\pm0.60)\times10^{-2}$&$$\\
$\hspace{0.15cm} 8.62$&$(1.06\pm0.45)\times10^{-2}$&$$\\
$\hspace{0.15cm} 8.74$&$(5.70\pm2.82)\times10^{-3}$&$$\\
$\hspace{0.15cm} 8.86$&$(1.41\pm0.32)\times10^{-2}$&$$\\
$\hspace{0.15cm} 8.98$&$(1.65\pm0.47)\times10^{-2}$&$$\\
$\hspace{0.15cm} 9.10$&$(8.70\pm3.30)\times10^{-3}$&$(1.62\pm0.63)\times10^{-2}$\\
$\hspace{0.15cm} 9.22$&$(1.44\pm0.39)\times10^{-2}$&$(1.71\pm0.80)\times10^{-2}$\\
$\hspace{0.15cm} 9.34$&$(1.85\pm0.35)\times10^{-2}$&$(1.14\pm0.37)\times10^{-2}$\\
$\hspace{0.15cm} 9.46$&$(1.55\pm0.20)\times10^{-2}$&$(1.79\pm0.43)\times10^{-2}$\\
$\hspace{0.15cm} 9.58$&$(1.60\pm0.16)\times10^{-2}$&$(1.97\pm0.33)\times10^{-2}$\\
$\hspace{0.15cm} 9.70$&$(1.74\pm0.15)\times10^{-2}$&$(1.69\pm0.36)\times10^{-2}$\\
$\hspace{0.15cm} 9.82$&$(1.83\pm0.11)\times10^{-2}$&$(1.70\pm0.09)\times10^{-2}$\\
$\hspace{0.15cm} 9.94$&$(1.71\pm0.09)\times10^{-2}$&$(1.80\pm0.17)\times10^{-2}$\\
$10.06$&$(1.64\pm0.09)\times10^{-2}$&$(1.87\pm0.17)\times10^{-2}$\\
$10.18$&$(1.56\pm0.07)\times10^{-2}$&$(1.79\pm0.16)\times10^{-2}$\\
$10.30$&$(1.45\pm0.07)\times10^{-2}$&$(1.64\pm0.10)\times10^{-2}$\\
$10.42$&$(1.28\pm0.07)\times10^{-2}$&$(1.64\pm0.09)\times10^{-2}$\\
$10.54$&$(1.01\pm0.06)\times10^{-2}$&$(1.46\pm0.09)\times10^{-2}$\\
$10.66$&$(6.62\pm0.33)\times10^{-3}$&$(1.36\pm0.08)\times10^{-2}$\\
$10.78$&$(4.02\pm0.19)\times10^{-3}$&$(1.14\pm0.07)\times10^{-2}$\\
$10.90$&$(2.11\pm0.12)\times10^{-3}$&$(7.97\pm0.41)\times10^{-3}$\\
$11.02$&$(9.11\pm0.55)\times10^{-4}$&$(4.99\pm0.33)\times10^{-3}$\\
$11.14$&$(3.49\pm0.32)\times10^{-4}$&$(2.69\pm0.13)\times10^{-3}$\\
$11.26$&$(9.17\pm1.23)\times10^{-5}$&$(1.34\pm0.09)\times10^{-3}$\\
$11.38$&$$&$(4.90\pm0.27)\times10^{-4}$\\
$11.50$&$$&$(1.66\pm0.17)\times10^{-4}$\\
$11.62$&$$&$(4.24\pm1.08)\times10^{-5}$\\
$11.74$&$$&$(1.68\pm0.66)\times10^{-5}$\\
\hline
\end{tabular}
\label{tab:schecltab}
\end{table}

With the galaxy total $J$- and $K_S$-band magnitudes defined as
described in the previous section, it is now possible to calculate the
galaxy luminosity function in these bands. This analysis extends that 
by Kochanek \etal (2001) and C01 by using over twice
as many galaxies, and also by adopting a different
treatment of the statistical errors. In addition, mock 2dFGRS
catalogues have been analysed to help determine the accuracy to which 
the luminosity functions can be recovered. The procedure for creating
these mock catalogues was described in Eke \etal (2004a), and
basically relies on populating dark matter haloes formed in the GIF
N-body simulation (Jenkins \etal 1998) using the semi-analytical
galaxy formation model described by Cole \etal (2000). In order to
match the 2dFGRS $\bj$ band luminosity function a luminosity-dependent
colour-preserving shift is applied to each galaxy. However, the
resulting model does
not match the near-IR galaxy luminosity function. Thus, for the
purposes of this paper, where the mock catalogue is being used to quantify
errors in the recovered quantities rather than testing the galaxy
formation model itself, an additional colour-preserving shift is
applied when considering near-IR luminosities or stellar masses. In
effect, each galaxy is made $15$ per cent brighter, this being the
factor necessary to increase the mean $K_S$ band luminosity density in
the model to agree with that in the real data. This simulation cube filled with
galaxies, hereafter referred to as the model, is then replicated and
observed with suitable flux limits, masks etc. (Norberg \etal 2002), 
creating a mock
catalogue. While the mean $K_S$ band luminosity density is the same in
mock and real catalogues, the luminosity function is not, so the
near-IR flux limits applied to the mock catalogue are chosen to
recover the same number of galaxies as are being considered from the
2MASS sample.

Figs~\ref{fig:schecj}
and~\ref{fig:scheck} show the $J$- and $K_S$-band luminosity functions
derived from both the mock catalogue and the 2dFGRS-2MASS data. Only
the results obtained using a stepwise
maximum likelihood estimator (SWML; Efstathiou, Ellis \& Peterson 1988)
are shown. A $1/V_{\rm max}$ estimator yields lower numbers of less
luminous galaxies because both mock and real observers happen to live in
underdense regions of universe. However, this estimator does give a
reliable abundance at higher luminosities, where it probes a larger
volume, and has thus been used to
fix the normalisation of the SWML curves. This was done using a
$\chi^2$ minimisation in either of the intervals
$10.1<\log_{10}[L_J/(h^{-2}L_\odot)]<11.2$ or
$10.4<\log_{10}[L_{K_S}/(h^{-2}L_\odot)]<11.4$. Errors on the recovered
SWML luminosity functions are determined from a jack-knife
procedure. This entails splitting the galaxy sample, by right
ascension, into ten subsamples containing equal surveyed areas, 
recalculating ten
different estimates ($n_{\i}$) of the luminosity function by 
rejecting each subsample in turn, then defining the uncertainty as
\begin{equation}
\sigma=\sqrt{\frac{10}{9}\sum_{\i=1}^{10} (n_{\i}-\bar{n})^2}, 
\end{equation}
where
($\bar{n}$) represents the luminosity function obtained using the full
sample. The $J$ and $K_S$ luminosity functions estimated from the mock
catalogue provide an accurate 
recovery of the underlying model function (shown with solid histograms),
although they do differ somewhat from the near-IR luminosity functions
estimated for the real data, despite having the same mean luminosity
densities. 
The deficit of high-luminosity galaxies in the mock catalogue represents an
interesting discrepancy between the semi-analytical model assumed in
the mock catalogue and the real Universe.

Having normalised the SWML luminosity function using the luminous end
of the $1/V_{\rm max}$ estimate, a $\chi^2$ fit can be performed
to determine the best-fitting Schechter (1976) function:
\begin{equation}
\phi(L)\d L=\phi_* \left(\frac{L}{L_*}\right)^\alpha {\rm exp}\left({-\frac{L}{L_*}}\right)
\frac{\d L}{L_*}.
\label{sch}
\end{equation}
In principle, the range over which this fit is performed need not be
restricted to the luminous galaxies. However, the impact 
of any potential low surface brightness incompleteness in 2MASS 
(Andreon 2002) is
likely to be apparent at lower luminosities, so the fit is restricted
to the same range of luminosities used to normalise the SWML function.
Thus, the derived parameters, which are listed in
Table~\ref{tab:schecl}, are also those that would have been found by
fitting to the $1/V_{\rm max}$-estimate of the luminosity
function.

The mean Universal luminosity densities implied
by the SWML-estimated luminosity functions are also included in
Table~\ref{tab:schecl}. 
Statistical uncertainties come from the dispersion in the
fits to the jack-knife samples. Note that, compared with the analysis
of C01, the uncertainties on the luminosity
function shape parameters ($L_*,\alpha$) are larger, because of the
limited range of luminosities over which the fit is
performed. In contrast, the statistical uncertainty on the
normalisation is lower because of the increased sample size (both
larger area and a deeper survey limit) and the
choice of normalisation method. Matching the abundance of high
luminosity galaxies is a better way of reducing the variance than
fitting to the counts, as C01 did. 

The statistical uncertainties are small compared to the
systematic uncertainties. These include imperfections in the sample, due to
incompleteness and misclassification in the parent 2dFGRS (Cross \etal
2004) and missing low surface brightness galaxies in the 2MASS
(Andreon 2002; Bell \etal 2003). It is this last effect that is most 
important. Both Andreon (2002) and Bell \etal
estimate that the 2MASS XSC misses $\sim 25$ per cent of the mean
luminosity density in the $K_S$ band. The mean near-IR luminosity densities in
Table~\ref{tab:schecl} are consistent with those of C01
and Bell \etal (2003), but have smaller statistical uncertainty.

\section{$K_S$-band group mass-to-light ratios}\label{sec:molk}

Following Eke \etal (2004b), who studied the $\bj$- and $\rf$-band
galactic content of groups of different size, the near-IR group
mass-to-light ratio ($\Upsilon_K$) is now considered. As the total
galaxy stellar mass correlates more strongly with the near-IR
flux than it does with the optical flux, $\Upsilon_K$ more effectively
reflects the efficiency with which stars have been formed in regions
that end up in these different sized haloes.

\subsection{Results for the model}\label{ssec:molmodk}

\begin{figure}
\centering
\centerline{\epsfxsize=8.6cm \epsfbox{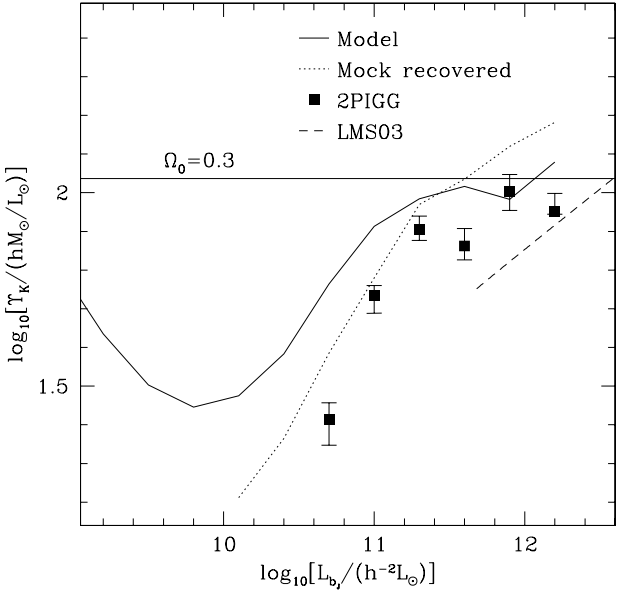}}
\caption{The variation of the mock and 2PIGG $K_S$-band mass-to-light
ratios with halo size, parametrized by the total $\bj$-band
luminosity. A horizontal line 
shows the mean mass-to-light ratio of the model, while the
solid curve traces the variation of the binned median
mass-to-light ratio in the model. The accuracy with which this is
recovered is shown by the dotted line, and the points display the
results obtained from the real 2PIGG sample. Only groups with no
neighbours within a distance of 
$d_{\rm min}/\Mpc=2+[\,10-\log_{10}(L_{b_{\rm J}}/\Lsol)]$ and out to
a maximum redshift of 
$z_{\rm max}=0.04+0.03\,[\,\log_{10}(L_{b_{\rm J}}/\Lsol)-10]$
are considered. Error bars show the 16th and 84th percentiles in each
bin. A dashed line traces the results of Lin \etal (2003), mapped onto
these axes using the median mass-to-$\bj$ light found by Eke \etal
(2004b), and assuming that the clusters have a concentration of $5$.
Taking the median group total mass-to-light ratio of Eke
\etal (2004b) maps the following luminosities 
$\log_{10}[L_{b_{\rm J}}/(h^{-2}L_\odot)]=(10,11,12)$ to masses
$\log_{10}[M_\odot/(h^{-1}M_\odot)]=(12.0,13.6,14.7)$.}
\label{fig:mollk}
\end{figure}

Fig.~\ref{fig:mollk} shows how the star formation efficiency varies
with halo size. The model used to populate the mock catalogues shows a
clear minimum in mass-to-light ratio at group sizes corresponding to
about one $L_*$ galaxy, as was seen in the optical bands (Eke \etal
2004b), as well as a
plateau for clusters with $\log_{10}(L_{b_{\rm J}}/\Lsol)>11.5$. The process
of identifying groups and inferring their mass-to-light ratios using a
dynamical mass estimator inevitably introduces errors. A dotted line
traces the mass-to-light ratio variation recovered from the mock catalogue.
In selecting the groups to be used to make this curve, only local and
isolated ones are taken, to avoid some sources of contamination. As a
consequence of these restrictions, which are detailed in the caption
of Fig.~\ref{fig:mollk}, the group mass-to-light ratio can be
recovered to within $\sim 50$ per cent for groups with
$\log_{10}(L_{b_{\rm J}}/\Lsol)>10.5$. The main source of bias is due to
interloping galaxies in the groups, which tend to increase the
luminosity more than the mass (the groups are found in redshift space
so interlopers have velocities similar to those of true group members).
An increase of $\Upsilon_K$ by a factor of $3$ in going from
$\log_{10}(L_{b_{\rm J}}/\Lsol)=10.5$ to $11.3$ is recovered, which is
comparable to that present in the model haloes, albeit shifted to
slightly larger halo luminosity.

\subsection{Observational data}\label{ssec:compmol}

The points in Fig.~\ref{fig:mollk}
represent the result of applying the same groupfinding,
mass-to-light measuring, and group selecting procedure to the 2dFGRS
as was applied to the mock catalogue. Again, a factor of $\sim 3$
increase in $\Upsilon$ is visible when going from the smaller groups to rich
clusters. All of this trend occurs at 
$\log_{10}(L_{b_{\rm J}}/\Lsol)<11.3$. 
Above $\log_{10}(L_{b_{\rm J}}/\Lsol) \approx 11.3$, the 
near-IR group mass-to-light ratio remains approximately constant. This
higher luminosity range corresponds to the range of cluster sizes in
the sample of Lin \etal (2003, LMS03). They find that the $K_S$ band
$\Upsilon_{500}\propto M_{500}^{0.31\pm0.09}$. This slope is steeper
than suggested by the data presented here. These are in better
agreement with the results of Kochanek \etal (2003) who find this
exponent to be $0.10\pm0.09$, despite going down to smaller groups
($\log_{10}(L_{b_{\rm J}}/\Lsol) \approx 10.8$) where the trend starts
to become apparent in the 2PIGG results.

For
the $110$ groups with $\log_{10}(L_{b_{\rm J}}/\Lsol)\ge11.3$, the median
$\Upsilon_K=89\pm3~h\Upsilon_\odot$ (statistical uncertainty only). 
If the 2PIGG mass-to-light ratios for such
clusters are overestimated by the same factor as in the mock
catalogues, then one should correct the median mass-to-light ratio
downwards accordingly to yield $\Upsilon_K=77\pm3~h\Upsilon_\odot$.
This is very similar to the value found by LMS03. 
For a sample of $19$ clusters with
$kT_X\ge3.7~$keV, they obtained $\Upsilon_K=76\pm4~h\Upsilon_\odot$ using
the 2MASS $K_S$-band group luminosities with a statistical 
background correction, and masses inferred from X-ray data and the
assumption of hydrostatic equilibrium. The LMS03 clusters extend to
blue luminosities of $\log_{10}(L_{b_{\rm J}}/\Lsol)\sim13$, off the
right hand side of Fig.~\ref{fig:mollk}, and it is these higher
luminosity clusters that bring the median mass-to-light ratio up into
agreement with the 2PIGG results. Kochanek \etal (2003) find
$\Upsilon_K=116\pm46~h\Upsilon_\odot$, which is consistent within the
statistical uncertainty. 

There are a number of potential systematic differences that blur
the comparison between the cluster $K_S$-band mass-to-light ratios
reported in different papers, at a level of $\sim 10$ per cent.
These include the possibility of galaxies not tracing dark
matter in real haloes as they do in the model universe and
incompleteness in the parent 2dFGRS. The LMS03
value is measured at a radius, $r_{500}$, enclosing a mean density 
of $500$ times the critical value, whereas the group definition used here
is tuned to the halo found by applying a friends-of-friends
group-finder, with a linking length equal to $0.2$ times the mean
interparticle separation, to a dark matter simulation. This algorithm
finds groups out to a radius $r_{\rm vir} \approx r_{200} > r_{500}$.
Rines \etal (2004) 
find that the variation of mass-to-light ratio with radius is slight
in their sample of 9 rich clusters. They use mass profiles
determined from the infall pattern of galaxies. Combining this with an
assumption that the dark matter haloes have Navarro,
Frenk \& White (NFW, 1997) profiles with a concentration of $5$, they
find that the mean mass-to-light ratios are
$\Upsilon(<r_{500})=78\pm7~h\Upsilon_\odot$ and
$\Upsilon(<r_{200})=88\pm9~h\Upsilon_\odot$. 

\subsection{Estimating $\Omega_{\rm m}$}\label{ssec:omega}

Given the $K_S$-band mass-to-light ratio of clusters, an assumption as
to how this compares to the mean Universal mass-to-light ratio and a
measurement of the mean luminosity density, yields an estimate of
$\Omega_{\rm m}$. If real clusters, like the model clusters, have
mass-to-light ratios that underestimate the mean Universal value, then
this exercise gives $\Omega_{\rm m}=0.21\pm0.01$ (statistical). This value
is obtained after applying two largely offsetting $\sim 10$ per cent
corrections, one assuming that the cluster mass-to-light ratio is
overestimated and the other that it provides an underestimate of the
universal value, so the statistical error is likely to bear little
relation to the true uncertainty on this estimate. It is interesting
to note that this estimate of $\Omega_{\rm m}$ is somewhat lower than that
produced by Eke \etal (2004b) using a similar method in the $\bj$ and
$\rf$ bands ($\Omega_{\rm m}\approx 0.27$). 

While the total $K_S$ band
luminosity density in the mock is the same as that in the real world,
the mock clusters have mass-to-light ratios that are $\sim 30$ per
cent greater than those in the real world. If the mock catalogues had
more faithfully represented the real shape of the $K_S$-band galaxy population,
then not only would more luminous galaxies have existed in the model but, 
perhaps, they would have preferentially occurred in rich
clusters. The net effect would have been to make the cluster
mass-to-light ratio a smaller fraction of the universal
mean. Consequently a larger correction upwards would have been required, and
when applied to the real data a higher $\Omega_{\rm m}$ would have
resulted. Given the systematic differences between the
mock catalogue and the real data in the $K_S$ band, it would be
premature to say that this low apparent value of $\Omega_{\rm m}$ represents
anything more than a challenge for semi-analytical models to place more
$K_S$-band luminosity into clusters.

\section{Stellar mass estimates}\label{sec:starsest}

The stellar mass of a galaxy is estimated from its broad band magnitudes
using simple stellar population synthesis models, as described at length
by C01. In summary, a grid of star formation histories is
constructed, in which both the e-folding time, $\tau$, which governs the star
formation rate, ($\psi \propto \exp(-t/\tau)$), and the metallicity of the
stars are varied. When combined with the assumption of a Kennicutt (1983)
stellar initial mass function (IMF) and the stellar population synthesis
models of Bruzual \& Charlot (2003), look-up tables of broad band
magnitudes versus redshift can be generated. Each observed set of galaxy
magnitudes is matched up with the theoretical model that best
reproduces these magnitudes at the redshift of the galaxy. The
theoretical track can then be used to estimate the $k+e$ correction that
should be applied to calculate the absolute magnitude of the galaxy (see
Section 2.2) and also to infer its mass-to-light ratio at $z=0$. There are
two major differences between the stellar masses computed here and those
inferred by C01. First $\rf$ magnitudes have now been
measured for the 2dFGRS objects. Second, 
models are selected from the theoretical grid to produce unique
predictions for colour versus redshift. This means, in practice, a
reduced dynamic range of the grid in e-folding time and
metallicity compared with that used by C01. The stellar metallicities
used here are now closer to solar.

With the availability of $\rf$ magnitudes, there is now more
information to help assess which theoretical model best matches the
colours of each observed galaxy. Throughout the rest of this paper,
two different stellar mass estimators will be referred to. The
$J$-band stellar mass will be that coming from using all four fluxes
($\bj$, $\rf$, $J$ and $K_S$) to choose a model star formation
history. The second estimate, the $\rf$-band inferred stellar mass, 
will be identical to the
$J$-band estimate when near-IR fluxes are available, but will also
include galaxies too faint to make the near-IR flux limit of the
sample. In these cases the $\rf$-inferred mass will be based on just
the $\bj$ and $\rf$ data. These two stellar mass estimates should thus
be considered to represent two different subsets of galaxies, with the
$\rf$-inferred stellar masses providing a deeper sample that reaches
higher redshifts and intrinsically less massive galaxies than the
$J$-band sample.

Note that the choice of stellar IMF introduces a large systematic
uncertainty in the recovered stellar masses. For instance, as C01
showed, a Salpeter IMF assigns almost twice as much stellar
mass per unit light as is inferred using the
Kennicutt IMF that is used throughout this paper. It should also be
stressed that the stellar masses referred to in this paper are those
currently locked up in stars. This differs from the total mass of star
formation by a factor that depends on how much stellar mass is
recycled. Following Cole \etal (2000), a recycled fraction of $R=0.42$
is adopted 
throughout this paper to approximate this effect for a Kennicutt IMF.

\begin{figure}
\centering
\centerline{\epsfxsize=10.8cm \epsfbox{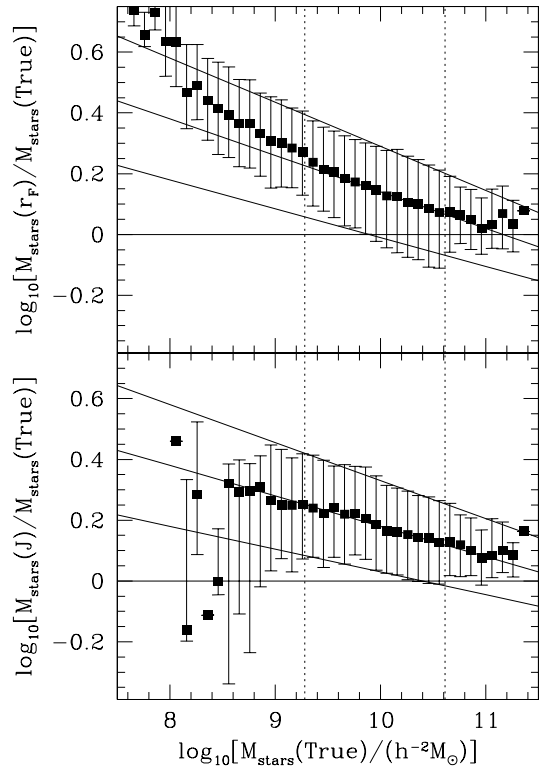}}
\caption{Errors in the inferred stellar masses using the $\rf$ (upper
panel) and $J$ band (lower panel) estimators.
The points show the median binned accuracy with which stellar
masses are recovered from the mock catalogue using the $\rf$-band stellar
masses. Error bars display the 16th
and 84th percentiles. All galaxies at $z<0.12$ have been used to
calculate these quantities. The solid lines represent the mean $\pm 1 \sigma$
errors used to model this bias in Section~\ref{sec:starst}. Dotted
vertical lines delimit the region containing the central $80$ per cent
of the stellar mass (\ie only $10$ per cent comes from both regions not
between dotted lines).}
\label{fig:mserr}
\end{figure}

Another uncertainty in the estimated stellar masses is the
appropriateness of the assumed exponential star formation history. It
is possible, however, to estimate the size of this error using the
mock catalogues. Fig.~\ref{fig:mserr} shows 
how well the recovery works for the $\rf$- and $J$-band stellar
masses referred to above, as a function of the true stellar mass in the model
galaxy. The inclusion of magnitude errors contributes little to the
range of stellar mass errors, which are instead dominated by the
choice of the grid of star formation histories. While the grid of
models assumes that the stars formed with a single exponential
timescale, the more realistic
semi-analytical galaxies undergo much more complex,
bursty histories. Consequently, some non-negligible error in the
estimated stellar masses is to be expected. Rough fits to the
typical errors in the stellar mass recovery and the scatter about this
error are shown with the lines in Fig.~\ref{fig:mserr}.
The equations of these lines are, for the $\rf$
band:
\begin{equation}
y=1.34-0.12x
\end{equation}
and
\begin{equation}
\sigma_y=0.4-0.025x,
\end{equation}
where $y=\log_{10}[M_{\rm stars}(\rm estimated)/M_{\rm stars}({\rm
true})]$ and $x=\log_{10}[M_{\rm stars}({\rm true})/(\hMsol)]$, 
and for the $J$ band
\begin{equation}
y=1.18-0.10x
\end{equation}
and the same $\sigma_y$ as above.

\section{The total stellar mass function}\label{sec:starst}

For each of the stellar mass estimates, it is possible to determine the
total stellar mass function in galaxies. The mock catalogues provide
guidance on the size of possible systematic errors arising from the
measurement procedure. Both $1/V_{\rm max}$ and bivariate stepwise
maximum likelihood (SWML; Sodr\'e \& Lahav 1993; Loveday 2000) 
methods have been employed. The former
method is sensitive to large scale fluctuations in the galaxy density
field and is thus less reliable for the low stellar masses that are
only seen locally. However, the high mass end 
with $10.4\le \log_{10}(M_{\rm stars}/\hMsol)\le 11.4$ provides a good
normalisation of the SWML method, which is insensitive to these
density fluctuations but only returns a shape for the stellar mass function.
A bivariate SWML method is necessary, with the variables
being the inferred stellar mass and the galaxy luminosity in the waveband
defining the survey (\ie $\bj$ or $J$ for the $\rf$- and $J$-band
stellar masses respectively).

\subsection{Systematic errors in the recovery}\label{ssec:systms}

\begin{figure}
\centering
\centerline{\epsfxsize=8.6cm \epsfbox{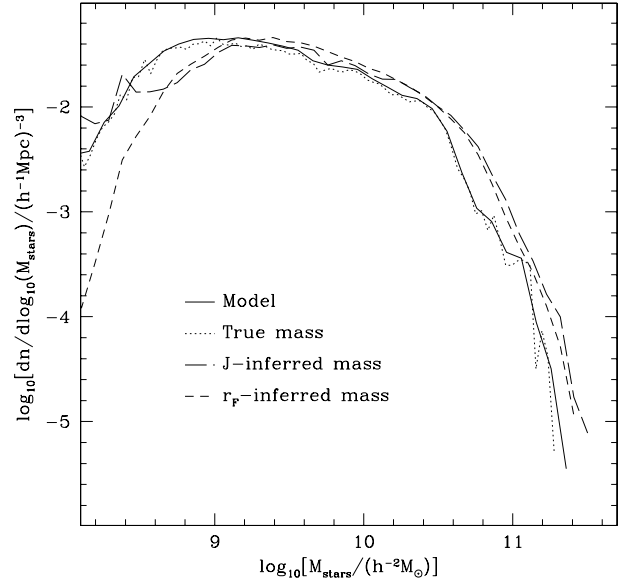}}
\caption{Systematic errors in the recovery of the global stellar mass
function. The solid line traces the stellar mass function in the
model. The dotted line shows a SWML estimate of the stellar mass
function from the mock catalogue using the true stellar masses, rather
than those inferred from the galaxy light. Long- and short-dashed lines
trace the normalised SWML stellar mass functions estimated from
the mock, using the $J$- and $\rf$-band-inferred stellar masses. The
abundance drops for galaxies with $\log_{10}(M_{\rm stars}/\hMsol)\lsim9$
because of the blue luminosity limit in the model galaxy population,
$M_{b_{\rm J}}\le-16$.}
\label{fig:schecmmod}
\end{figure}
Fig.~\ref{fig:schecmmod} shows how well the stellar mass function can be
recovered from the mock catalogue. Only galaxies with $z<0.12$ are
used in constructing these functions. It is immediately apparent that the
effect of the measurement errors in the galaxy stellar masses is to
produce an overestimate of the abundance of galaxies at the higher stellar
masses. This, in turn, is reflected in an overestimation of the contribution
to the cosmic density from stars residing in galaxies. While the
model value is $\Omega_{\rm stars}h=1.12 \times 10^{-3}$ (from
integrating under the solid black line multiplied by $M_{\rm stars}$), 
and the fraction of universe that is actually probed by the mock
survey only has $\Omega_{\rm stars}h=1.03 \times 10^{-3}$ (dotted
line) because of variations due to large scale structure, the
recovered SWML values are $\Omega_{\rm stars}h=1.46 \times
10^{-3}$ (short-dashed line) and $\Omega_{\rm stars}h=1.45 \times
10^{-3}$ (long-dashed line) for the $\rf$- and $J$-band measurements
respectively. To give an idea of the size of the error that can be
introduced by using the $1/V_{\rm max}$ stellar mass function, rather
than the suitably normalised SWML one, the $J$-band $1/V_{\rm max}$
stellar mass function yields a value of 
$\Omega_{\rm stars}h=1.26 \times 10^{-3}$, $\sim 15$ per cent down on the
SWML estimate, which is unaffected by the large-scale density fluctuations.

\begin{figure}
\centering
\centerline{\epsfxsize=8.6cm \epsfbox{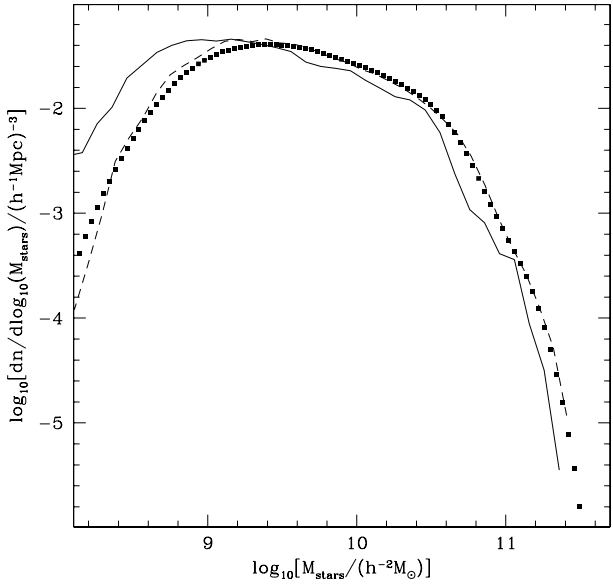}}
\caption{How stellar mass uncertainties affect the estimated stellar
mass function. The solid line traces the stellar mass function in the
model. The dashed line shows the SWML estimate of the stellar mass
function from the mock catalogue using the $\rf$-band stellar masses.
Points show the effect of convolving the model stellar mass function
with a mass-dependent Gaussian, of mean and width given by the
straight lines in Fig.~\ref{fig:mserr}.}
\label{fig:msconv}
\end{figure}

Fig.~\ref{fig:msconv} shows the stellar mass function obtained by
convolving the model with a Gaussian whose mean and
width varies with stellar mass according to the lines in
Fig.~\ref{fig:mserr}. The fact that this matches very well the
stellar mass function inferred from the $\rf$-band mock catalogue
demonstrates that this is the uncertainty that causes the
systematic bias in the recovery of the stellar mass function. The
$J$-band result is very similar.

From the mock catalogue, it is apparent that a SWML estimator of
the stellar mass function is preferable to $1/V_{\rm max}$, and
that the large uncertainties in 
the stellar mass estimation lead to an overestimate of $\Omega_{\rm
stars}h$ by $\sim 40$ per cent. Given that this overestimation
results largely from the method chosen to assign stellar masses, it
seems likely that such a systematic error would also be present when
this technique is applied to real data. While the size of this error
is merely comparable with the systematic uncertainty arising from the
choice of stellar IMF, it is larger than the statistical uncertainties
quoted by C01 and Bell \etal (2003), and should be
corrected for. Note that Bell \etal use a different procedure to calculate
stellar masses, so the bias introduced by their method may not be
quite the same as that found in this work.

\subsection{Comparison with other studies}\label{ssec:compms}

\begin{table*}
\begin{center}
\caption{Best-fitting Schechter function parameters for the different
galaxy SWML stellar mass functions. The SWML estimates are normalised
using a $\chi^2$ fit to the $1/V_{\rm max}$-estimated abundances
over the range $10.4\le \log_{10}(M_{\rm stars}/\hMsol)\le 11.4$. A
$\chi^2$ fit is then performed to determine the best-fitting Schechter
functions. This latter fit is carried out over larger ranges of
stellar masses. For the $J$-inferred case, galaxies within $10.0\le
\log_{10}(M_{\rm stars}/\hMsol)\le 11.4$ are used, and this is
extended to $9.5\le \log_{10}(M_{\rm stars}/\hMsol)\le 11.4$ for the
$\rf$-inferred function. 
Also listed are the mean Universal stellar mass densities derived
from the SWML functions. Only statistical uncertainties are quoted.
More important systematic uncertainties are discussed in the
text. The final column contains the corrected estimates of the mean
stellar density, taking into account the overestimation described in
Section~\ref{sec:starsest}. Note that these stellar masses are all
derived under the assumption that a Kennicutt IMF is universally applicable.}
\begin{tabular}{llllll} \hline
band & $~~~~M_*/(h^{-2}M_\odot)$ & ~~~~~~~~$\alpha$ &
 ~~~$\phi_*/(h^{-1}{\rm Mpc})^{-3}$ & $\Omega_{\rm stars}h$ & $\Omega_{\rm stars,c}h$ \\
 \hline
$M_{\rm stars}(\rf)$&$(3.28\pm0.24)\times10^{10}$&$-0.99\pm0.05$&$(1.20\pm0.12)\times10^{-2}$&$(1.48\pm0.01)\times10^{-3}$&$(1.04\pm0.01)\times10^{-3}$\\
$M_{\rm stars}(J)$&$(3.70\pm0.26)\times10^{10}$&$-0.93\pm0.07$&$(1.01\pm0.09)\times10^{-2}$&$(1.31\pm0.04)\times10^{-3}$&$(0.93\pm0.03)\times10^{-3}$\\
\hline
\end{tabular}
\label{tab:schecm}
\end{center}
\end{table*}
\begin{table}
\caption{The recovered stellar mass functions from the real data,
uncorrected for the measurement bias described in the text. The units
of stellar mass are $\hMsol$ and the quoted abundances are
$\d n/\d\log_{10}(M_{\rm stars})$ in $(h^{-1}{\rm Mpc})^{-3}$. To correct for
the bias introduced by the uncertainties in estimating stellar masses,
applying a global shift of $-0.15$ in $\log_{10}(M_{\rm stars})$ is a
good approximation.}
\begin{tabular}{lll} \hline
$\log_{10}(M_{\rm stars})$ & ~~~~~~~$\rf$ band & ~~~~~~~$J$ band \\
\hline
$\hspace{0.4cm} 9.0$&$(3.90\pm0.32)\times10^{-3}$&$(1.21\pm0.30)\times10^{-2}$\\
$\hspace{0.4cm} 9.1$&$(3.50\pm0.26)\times10^{-2}$&$(1.32\pm0.44)\times10^{-2}$\\
$\hspace{0.4cm} 9.2$&$(3.16\pm0.22)\times10^{-2}$&$(1.42\pm0.24)\times10^{-2}$\\
$\hspace{0.4cm} 9.3$&$(2.90\pm0.21)\times10^{-2}$&$(1.43\pm0.26)\times10^{-2}$\\
$\hspace{0.4cm} 9.4$&$(2.67\pm0.21)\times10^{-2}$&$(1.57\pm0.28)\times10^{-2}$\\
$\hspace{0.4cm} 9.5$&$(2.47\pm0.18)\times10^{-2}$&$(1.79\pm0.20)\times10^{-2}$\\
$\hspace{0.4cm} 9.6$&$(2.34\pm0.15)\times10^{-2}$&$(1.72\pm0.14)\times10^{-2}$\\
$\hspace{0.4cm} 9.7$&$(2.16\pm0.16)\times10^{-2}$&$(1.60\pm0.13)\times10^{-2}$\\
$\hspace{0.4cm} 9.8$&$(2.09\pm0.14)\times10^{-2}$&$(1.76\pm0.15)\times10^{-2}$\\
$\hspace{0.4cm} 9.9$&$(1.94\pm0.15)\times10^{-2}$&$(1.77\pm0.16)\times10^{-2}$\\
$\hspace{0.4cm} 10.0$&$(1.92\pm0.15)\times10^{-2}$&$(1.48\pm0.11)\times10^{-2}$\\
$\hspace{0.4cm} 10.1$&$(1.74\pm0.13)\times10^{-2}$&$(1.56\pm0.12)\times10^{-2}$\\
$\hspace{0.4cm} 10.2$&$(1.66\pm0.13)\times10^{-2}$&$(1.50\pm0.09)\times10^{-2}$\\
$\hspace{0.4cm} 10.3$&$(1.53\pm0.11)\times10^{-2}$&$(1.37\pm0.09)\times10^{-2}$\\
$\hspace{0.4cm} 10.4$&$(1.33\pm0.10)\times10^{-2}$&$(1.27\pm0.09)\times10^{-2}$\\
$\hspace{0.4cm} 10.5$&$(1.17\pm0.10)\times10^{-2}$&$(1.01\pm0.05)\times10^{-2}$\\
$\hspace{0.4cm} 10.6$&$(8.54\pm0.62)\times10^{-3}$&$(8.87\pm0.64)\times10^{-3}$\\
$\hspace{0.4cm} 10.7$&$(6.32\pm0.45)\times10^{-3}$&$(5.73\pm0.34)\times10^{-3}$\\
$\hspace{0.4cm} 10.8$&$(4.50\pm0.33)\times10^{-3}$&$(4.29\pm0.25)\times10^{-3}$\\
$\hspace{0.4cm} 10.9$&$(2.42\pm0.18)\times10^{-3}$&$(3.00\pm0.21)\times10^{-3}$\\
$\hspace{0.4cm} 11.0$&$(1.25\pm0.10)\times10^{-3}$&$(1.65\pm0.10)\times10^{-3}$\\
$\hspace{0.4cm} 11.1$&$(5.63\pm0.51)\times10^{-4}$&$(8.31\pm0.71)\times10^{-4}$\\
$\hspace{0.4cm} 11.2$&$(3.09\pm0.36)\times10^{-4}$&$(4.19\pm0.41)\times10^{-4}$\\
$\hspace{0.4cm} 11.3$&$(1.45\pm0.23)\times10^{-4}$&$(1.71\pm0.20)\times10^{-4}$\\
$\hspace{0.4cm} 11.4$&$(3.89\pm0.97)\times10^{-5}$&$(8.87\pm2.25)\times10^{-5}$\\
$\hspace{0.4cm} 11.5$&$(1.60\pm0.77)\times10^{-5}$&$(1.66\pm1.01)\times10^{-5}$\\
\hline
\end{tabular}
\label{tab:schecmtab}
\end{table}

\begin{figure}
\centering
\centerline{\epsfxsize=8.6cm \epsfbox{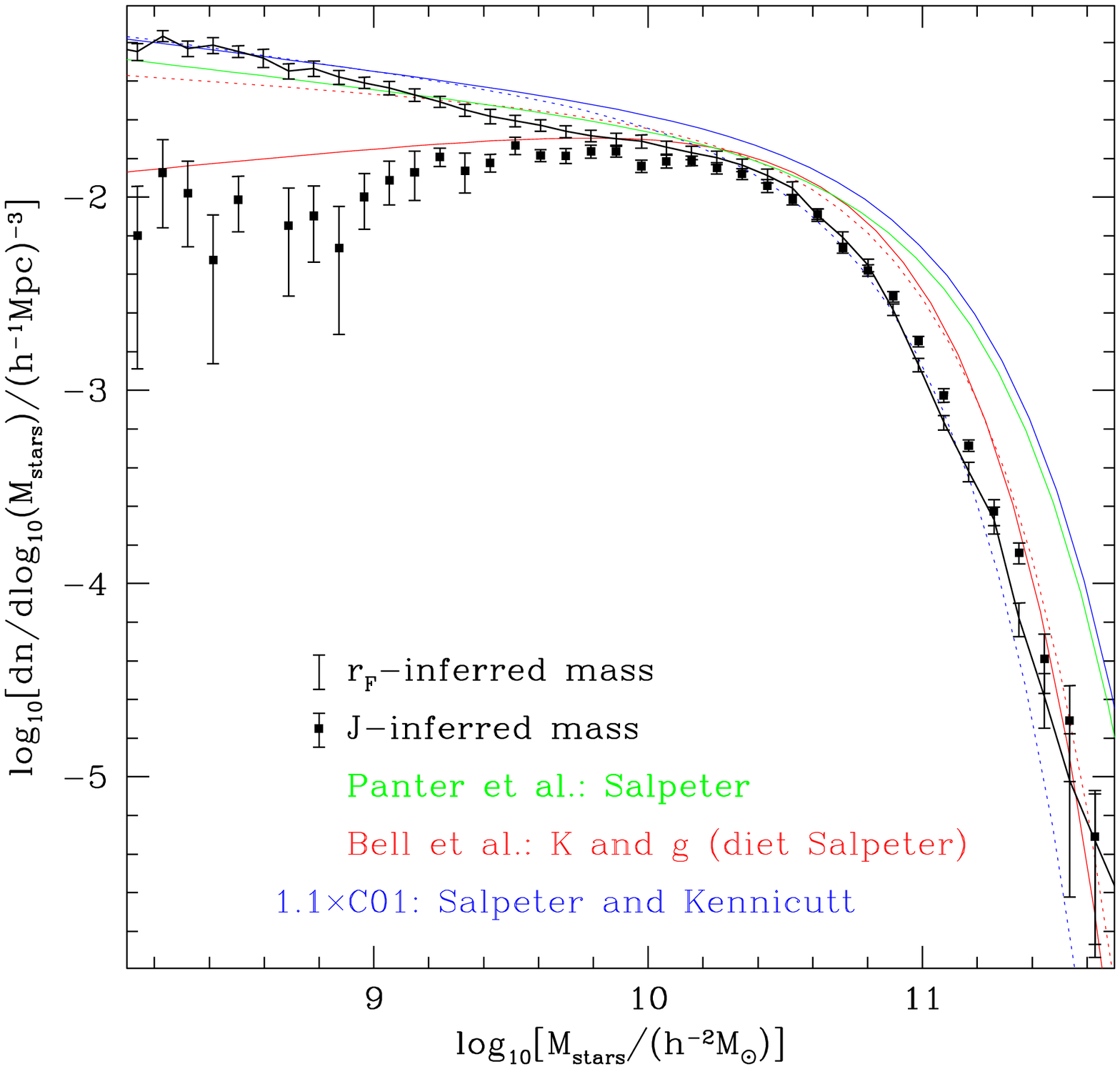}}
\caption{The uncorrected stellar mass functions calculated using the real
2dFGRS-2MASS sample of galaxies. SWML results from both $J$- (filled squares)
and $\rf$-band (black line with errors) stellar mass estimators are
shown. Also shown are the best-fitting Schechter 
functions advocated by Panter, Heavens \& Jimenez (Salpeter IMF; green);
C01 (Salpeter; solid blue); C01 (Kennicutt; dotted blue); and Bell
\etal - $K$-band (solid red) and $g$-band (dotted red).}
\label{fig:schecmreal}
\end{figure}
Applying the $1/V_{\rm max}$ and SWML estimators to the full
2dFGRS-2MASS sample within $z=0.12$ yields the results shown in
Fig.~\ref{fig:schecmreal}. Comparing with the results of C01, who
used both Kennicutt (1983) and Salpeter (1955) IMFs, it is apparent that
these new results are concordant at higher stellar masses.
For the smaller systems, the $J$-band mass function, which should
be directly comparable with that of C01, shows a deficit.
The $\rf$-band function does include most of the low stellar
mass systems found by C01. Note that the stellar masses of the
best-fitting Schechter function of C01 have been multiplied by
$1.1$ to take into account, roughly, the difference between Kron and
total magnitudes. The discrepancy with C01 
is mainly for the low-mass galaxies, and thus does not greatly impact
upon the inferred value of $\Omega_{\rm stars}h$. Assuming a Kennicutt
IMF, C01 found $\Omega_{\rm stars}h=(1.6\pm0.24)\times10^{-3}$,
whereas the corresponding numbers found here are 
$(1.48\pm0.01)\times10^{-3}$ and
$(1.31\pm0.04)\times10^{-3}$ for the $\rf$- and $J$-band respectively. As
described above, it is entirely conceivable that these, already lower,
estimates of $\Omega_{\rm stars}h$ are nevertheless still
overestimated by $\sim 40$ per cent as a result of the errors inherent
in assigning stellar masses to galaxies. Correcting these estimates under
the assumption that they suffer the same fractional overestimates as 
the mock data, gives the corrected $\Omega_{\rm stars}h$ values in the
final column of Table~\ref{tab:schecm}. Also listed in
Table~\ref{tab:schecm} are the
best-fitting Schechter function parameters obtained by a $\chi^2$ fit to
the normalised SWML stellar mass functions over the range
$9.5 \le \log_{10}(M_{\rm stars}/\hMsol)\le 11.4$ for the $\rf$ band
stellar masses and $10.0 \le \log_{10}(M_{\rm stars}/\hMsol)\le 11.4$
for the $J$ band stellar masses. The quoted statistical uncertainties
are estimated from the scatter in the fits to the jack-knife subsamples. In
addition to the systematic uncertainties in the luminosity
function estimation, a further source of systematic uncertainty is the
poorly known stellar IMF. To give a rough idea of how large an effect this can
have, C01 find that a Salpeter IMF yields a mean stellar mass density
that is almost twice as large as that implied by the Kennicutt IMF.

The difference between the stellar mass
functions inferred using the two different mass estimators
is consistent with what was found by Bell \etal -- namely that the sample
selected by 2MASS flux is incomplete as a result of missing low
surface brightness galaxies. This was shown by Bell \etal comparing a
2MASS $K$-selected sample with a SDSS $g$-band one. Their results
exhibit a deficit of 2MASS-selected galaxies at $\log_{10}[M_{\rm
stars}/(h^{-2}M_\odot)]<10$, very similar to the mass at which the
$J$- and $\rf$-band results found here begin to differ.

Combining the estimate of the mean stellar density and the luminosity
density from the previous section yields an estimate of the mean
stellar mass-to-light ratio. Taking the average of the uncorrected
$\Omega_{\rm stars}(\rf)$ and $\Omega_{\rm stars}(J)$ values, because
it is unclear which is better, gives a mean $K_S$ band stellar 
mass-to-light ratio of $0.55 \Upsilon_\odot$. If the corrected stellar
mass densities are employed instead, then this reduces to $0.39
\Upsilon_\odot$. 
Even the uncorrected value is $\sim 30$ per cent lower than that found 
by C01 ($0.73\pm0.15 \Upsilon_\odot$). 

\section{Stellar masses in different sized groups}\label{sec:starsg}

\begin{figure}
\centering
\centerline{\epsfxsize=8.6cm \epsfbox{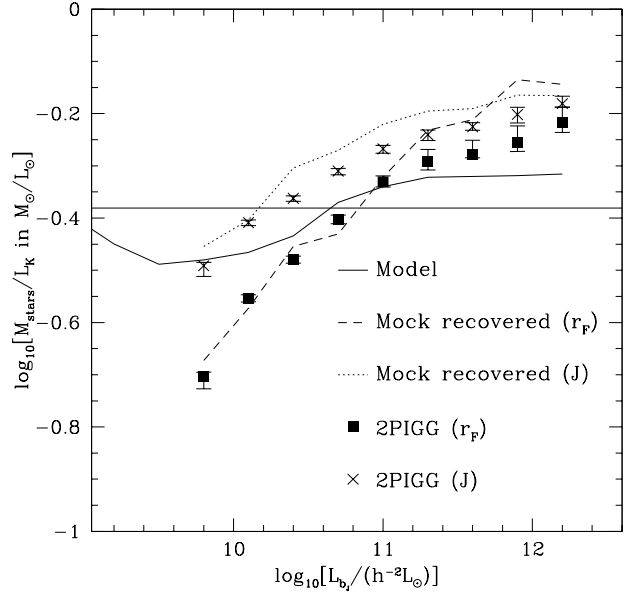}}
\caption{Stellar mass-to-light ratios ($K_S$ band) in the groups. The
horizontal line shows the mean value in the model, and the solid curve
shows the variation of the median with group size present in the model.
Dashed and dotted lines trace the results recovered from
the mock catalogues using $\rf$- and $J$-inferred stellar masses
respectively. The corresponding median values in the 2PIGG sample are
shown with filled squares and crosses.
Taking the median group total mass-to-light ratio of Eke
\etal (2004b) maps the following luminosities 
$\log_{10}[L_{b_{\rm J}}/(h^{-2}L_\odot)]=(10,11,12)$ to masses
$\log_{10}[M_\odot/(h^{-1}M_\odot)]=(12.0,13.6,14.7)$.}
\label{fig:msolk}
\end{figure}

The 2PIGG catalogue allows a detailed study of
how the galaxy stellar mass is distributed among groups of
different size. This Section presents measurements
of the stellar mass-to-light ratio and the stellar mass fraction, both
as functions of group size.

When calculating total group stellar masses, it is necessary to
include that contributed by group members that do not make it
into the flux-limited sample. This is done through the global bivariate SWML
distribution of stellar mass and galaxy luminosity. Integrating this
distribution over all stellar masses gives the fraction of stellar
mass above a particular galaxy luminosity. The reciprocal of this
fraction is the factor by which the total observed stellar mass must be
increased in order to obtain the total group stellar mass.

\subsection{Group stellar mass-to-light ratios}\label{ssec:gsmol}

The stellar mass-to-light ratio in the $K_S$ band is shown as a function of
total group $\bj$ luminosity in Fig.~\ref{fig:msolk}. Only groups with
$z<z_{\rm max}=0.05+0.02[\,\log_{10}(L_{b_{\rm J}}/\Lsol)-10]$ are
used, this being an appropriate compromise between having enough
groups, and being able to see most of their stellar mass and luminosity.
It is apparent that the values recovered from the mock generally
overestimate the model mass-to-light ratio, particularly for the
$J$-inferred stellar masses. This can be traced directly to the
results shown in Figure~\ref{fig:mserr}. The
slightly greater overestimation in stellar masses at the high mass end
for the $J$-band estimates is sufficient to account for this
difference. A steeper trend is seen in the $\rf$-inferred stellar
mass-to-light ratio than in the $J$-band one. This is because for the
less luminous groups, the $\rf$ measurements include groups with no
galaxies that are detected in the near infra-red, and these groups
preferentially have their stellar masses underestimated, thus dragging down
the median $\rf$-inferred stellar mass-to-light ratios.

Very similar trends are apparent in the real 2PIGG data, with a
small offset to lower mass-to-light ratios relative to the mock. 
For the largest 2PIGGs, the stellar mass-to-light ratio is $\sim
0.63 \Upsilon_\odot$. However, if one assumes that this is
overestimated by the same 
factor as in the mock catalogue, then one should expect that
the underlying stellar mass-to-light ratio in the most luminous 2PIGGs
should be $ \sim 0.45 \Upsilon_\odot$.

Even with the overestimation of the stellar masses, the
inferred stellar mass-to-light ratios are still lower than
was assumed by LMS03 ($\sim 0.8 \Upsilon_\odot$
for the richest clusters). This value was derived from dynamical data
coupled to a maximum stellar mass model, so it represents an upper
limit on the stellar mass-to-light ratio. Alternatively, if the
Kennicutt IMF were changed to one producing more stellar mass at a
given luminosity (for instance the `diet' Salpeter IMF of Bell \& de
Jong 2001), then the stellar mass-to-light ratio found here could be
increased substantially to agree with that assumed by LMS03.

\begin{figure}
\centering
\centerline{\epsfxsize=10.6cm \epsfbox{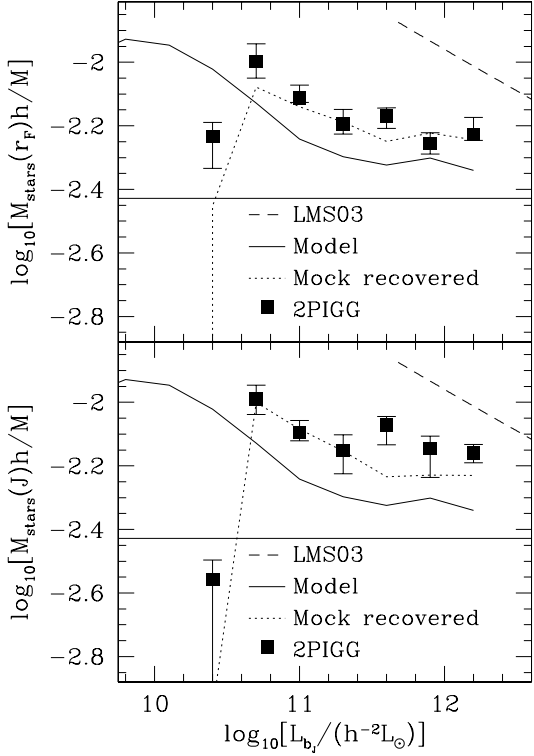}}
\caption{The ratio of stellar to total dynamical mass inferred using
the $\rf$ (upper panel) and $J$ (lower panel) band estimators. A
horizontal line
shows the ratio $\Omega_{\rm stars}/\Omega_{\rm m}$ in the model, and the
solid curve traces the variation of the median stellar mass
fraction with group size in the model according to the semi-analytical 
prediction. A dotted line follows the results recovered from the mock
catalogue, and the points show the median behaviour found in the 2PIGG
sample. The dashed line traces the results found by LMS03.}
\label{fig:moms}
\end{figure}

\begin{figure}
\centering
\centerline{\epsfxsize=8.6cm \epsfbox{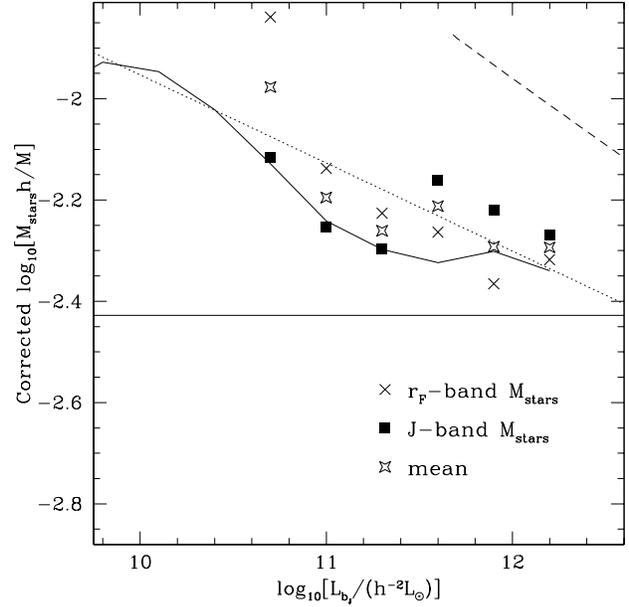}}
\caption{Corrected stellar mass fractions in the real data (symbols),
compared with the model (solid line). The dotted line is a
least-squares fit to the stars), and the dashed line shows the results
of LMS03.}
\label{fig:momscor}
\end{figure}

\subsection{Group stellar mass fractions}\label{ssec:gsmom}

Fig.~\ref{fig:moms} shows how the group stellar mass
fractions (\ie the mass currently in stars divided by the total group
mass) depend on group size for the $\rf$- and $J$-inferred stellar
masses. The horizontal line shows the mean stellar mass fraction in
the model universe, and the other solid line traces the variation in
stellar mass fraction with group size in the model, 
parametrised through the total
group $\bj$ band luminosity. The median values recovered from the
mock catalogues are shown with a dotted line. In constructing this
curve, only groups with $z<z_{\rm max}=0.02+0.04[\,\log_{10}(L_{b_{\rm
J}}/\Lsol)-10]$ are used. About $100$ groups contribute to
the bin with $\log_{10}[L_{b_{\rm J}}/(\Lsol)]=11$ and $\sim 25$
groups to the bin at $\log_{10}[L_{b_{\rm J}}/(\Lsol)]=11.9$.
For groups with
$\log_{10}[L_{b_{\rm J}}/(\Lsol)]\ge 10.7$ the variation of stellar
mass fraction is recovered well, albeit with a $\sim 25$ per cent
offset resulting largely from the overestimation of stellar masses.
This trend of increasing stellar mass fraction with
decreasing group size is apparent in both the model and the mock catalogue. 
The 2PIGG data points show a generally similar behaviour to that in the 
mock catalogue. As for the case where stellar masses are inferred from
the $\rf$ band, the $J$-band stellar mass estimator yields a
trend of increasing stellar mass fraction with decreasing group
size for the data. 

The dashed lines in Fig.~\ref{fig:moms} show the results of LMS03.
They measure the mass fraction within a
smaller radius than is used to define groups here. Thus, to calculate
where their results fit on these plots, it has been necessary to
assume a conversion from their $M_{500}$ to virial mass, 
$M_{\rm vir}$, (this is
taken simply as a factor $1.9$ for a typical cluster dark matter
density profile with a concentration, $c_\Delta=5$), to use the
typical mass-to-$\bj$ band light as a function of group size found by
Eke \etal (2004b), and to account for their use of $h_{70}$. For the
largest clusters, they find a stellar mass fraction of $\sim
9\times10^{-3}h^{-1}$, in contrast to the $\sim
6.3\times10^{-3}h^{-1}$ found here. This is similar to the difference
between the stellar mass-to-light ratios in the two studies.

Interestingly, the variation of stellar mass fraction
with cluster size found by LMS03 exceeds that found here, despite
their clusters being in a region where no significant trend is
apparent in the 2PIGG results. This is reminiscent of the situation
with the total mass-to-light ratio comparison. In fact, the 2PIGG
results would show only weak evidence of any trend if it were not for the
data at $\log_{10}[L_{b_{\rm J}}/(\Lsol)]=10.7$. One could envisage
that a strong trend of decreasing stellar mass fraction with increasing
total mass could be artificially created if the lowest mass clusters
preferentially had their masses underestimated.

Under the assumption that the real mass fractions are
overestimated in the same way as those in the mock catalogue, a
corrected stellar mass fraction can be
derived. Fig.~\ref{fig:momscor} shows a comparison between the model
from which the mock was created and the corrected results for the real
data. The 
different symbols show the two different stellar mass estimates and
the mean (in log space) of these two. A dashed line traces the
least-squares fit to the mean values. This line has the equation:
\begin{equation}
\log_{10}(M_{\rm stars}h/M)=-0.38-0.16~\log_{10}[L_{b_{\rm J}}/(\Lsol)].
\label{moms}
\end{equation}
For the biggest clusters, the typical corrected stellar mass fraction is only
$\sim 5\times10^{-3}h^{-1}$, about $25$ per cent lower than the raw value.

\section{Where are the stars?}\label{sec:where}
\begin{figure*}
\centering
\centerline{\epsfxsize=19cm \epsfbox{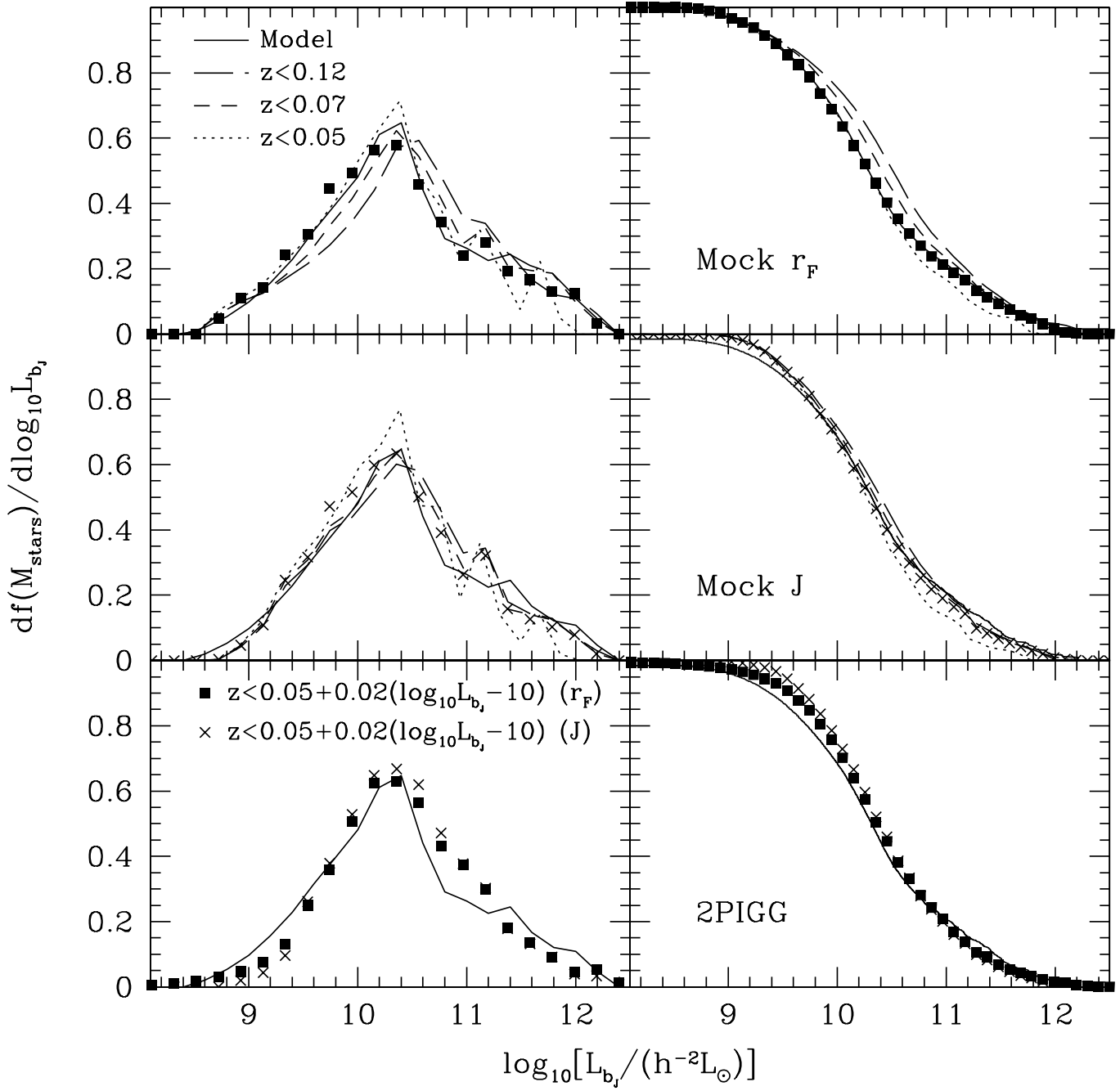}}
\caption{The fraction of all stellar mass within groups as a function of
group size, parametrized by group $\bj$-band luminosity. Differential
and the corresponding cumulative distributions are shown in the left
and right columns respectively. The solid 
line in each panel shows the model distribution, while the other
curves in the top two rows trace the estimates recovered as the maximum
redshift is changed from $0.05$ (dotted) to $0.07$ (short dashed) and
$0.12$ (long dashed). Results of using $\rf$-inferred stellar masses
in the mock catalogue are shown in the top row, whereas the middle
row contains the $J$-inferred stellar masses. The symbols show the
estimates with a variable upper redshift limit, dependent on the total
group $\bj$ band luminosity, as quantified in the legend.
The bottom row shows the 2PIGG results, with the variable upper
redshift limits for both the $\rf$- and $J$-inferred stellar mass
estimates, as well as the reference model.
Taking the median group total mass-to-light ratio of Eke
\etal (2004b) maps the following luminosities 
$\log_{10}[L_{b_{\rm J}}/(h^{-2}L_\odot)]=(10,11,12)$ to masses
$\log_{10}[M_\odot/(h^{-1}M_\odot)]=(12.0,13.6,14.7)$.}
\label{fig:trivmod}
\end{figure*}

Having determined both the total stellar mass function of galaxies 
and the variation of stellar mass content with group size, the next
natural step is to determine how all stars are partitioned among
groups. It should be noted that this work only considers stars that
are associated with galaxies, ignoring any population of
intergalactic stars. In large galaxy clusters, a few tens of per cent
of the total cluster stellar mass may reside in intergalactic stars
(\eg Zibetti \& White 2004; Lin \& Mohr 2004). However, only a
small fraction of all stars reside in such large clusters, as
will be shown. 

To determine the fraction of stars in groups of
a given luminosity, a trivariate SWML method has been used. By
choosing group luminosity to describe the group size, rather than
mass, it is possible to use groups containing only a single
galaxy. These contribute a significant fraction of the galaxies, and
hence stellar mass, and are thus crucial in forming a full picture of
where all of the stars reside. The trivariate SWML method
extends the method used in the previous sections by adding an extra
variable, namely the luminosity of the group in which any particular
galaxy exists. Following the nomenclature of Loveday (2000), the
space density of galaxies, $\phi(L_\g,L,M)$, with luminosity $L$,
stellar mass $M$ and residing in a group with total $\bj$ luminosity 
$L_\g$, can be used to write the probability of galaxy i taking
particular values as
\begin{equation}
p_i=\frac{\phi(L^\i_\g,L^\i,M^\i)}{\int_{L_{\rm g,min}^\i}^\infty\int_{L_{\rm
min}^\i}^{L_{\rm max}^\i}\int_0^\infty \phi(L_\g,L,M)~ \d L_\g \d L \d M}.
\label{trivprob}
\end{equation}
The lower
integration limit for the group total $\bj$ luminosity takes account
of the fraction of group luminosity that lies in galaxies beneath the
flux limit of the 2dFGRS, assuming that group galaxies follow the
Schechter function determined by Norberg \etal (2002) for the
entire population (as was done by Eke \etal 2004b when defining group
luminosity). The galaxy luminosity and stellar mass variables are
either $\bj$-band luminosity and $\rf$-band stellar mass or $J$-band
luminosity and stellar mass. As shown by the integration limits, all
stellar masses are accessible at any given galaxy luminosity -- this
was also assumed for the bivariate case earlier. Performing the usual
operations to maximise this probability while describing the galaxy
space density in a stepwise manner leads to the following estimate of
the density of galaxies residing in groups with
\begin{equation}
L_\g^\j-\Delta L_\g/2 < L_\g \le L_\g^\j+\Delta L_\g/2,
\end{equation}
having luminosity
\begin{equation}
L^\k-\Delta L/2 < L \le L^\k+\Delta L/2,
\end{equation}
and stellar mass
\begin{equation}
M^{\l}-\Delta M/2 < M \le M^{\l}+\Delta M/2:
\end{equation}
\begin{equation}
\phi_{\j\k\l}=\frac{n_{\j\k\l}} {\sum_{i=1}^{N} \left[
\frac{H_{\i\j\k\l}\Delta L_\g \Delta L \Delta M}{\sum_{\p=1}^{N_{L_\g}}\sum_{\q=1}^{N_L}\sum_{\r=1}^{N_M}\phi_{\p\q\r}H_{\i\p\q\r}\Delta L_\g \Delta L \Delta M}\right]}.
\end{equation}
$n_{\j\k\l}$ represents the total number of galaxies in bin
$(\j,\k,\l)$ and $H_{\i\p\q\r}$ is a trivariate version of the ramp
function as described by Loveday (2000). $N_{L_\g}$, $N_L$ and $N_M$
are the numbers of bins in group luminosity, galaxy luminosity and
galaxy stellar mass respectively.

The trivariate function can be projected to recover the galaxy
luminosity function or the stellar mass function, for
example. However, by projecting along the individual galaxy luminosity
and stellar mass directions, the distribution of stars among groups
of different luminosity can be retrieved.

Fig.~\ref{fig:trivmod} shows both the differential and cumulative
distributions of stellar mass among different sized groups. While the
distributions recovered from the mock catalogue are quite 
similar to that present in the model, there is a trend in the mock of
placing a higher fraction of the stars in larger groups as the upper
redshift of the sample is increased. This is particularly apparent for
the $\rf$-band results, because the mock $J$-band sample is also restricted
by the relatively high 2MASS flux limit. For the $z<0.05$ samples, a
lack of large local clusters in the mock catalogue means that the results
significantly underestimate the fraction of all stars in groups with
$\log_{10}(L_{b_{\rm j}})>11$. As the upper redshift limit is increased, the
volume encloses a fairer sample of the mock universe. Furthermore, the
correction for missing group luminosity increases, which leads to an
overestimate of the abundance of the more luminous groups.
Consequently, the stellar mass is too frequently assigned to larger
groups than should be the case. To correct for these systematic
biases, one can choose an upper redshift that varies with group
luminosity. The points in Fig.~\ref{fig:trivmod} show how the redshift
limit can be chosen so that the model distribution is very accurately
recovered. When the cut on group luminosity at a given redshift is
introduced, it is necessary to alter the lower limit for the group
luminosity integration in eq.~(\ref{trivprob}).

Assuming that similar systematic biases are appropriate in both the
mock and real data, the points in the bottom row of
Fig.~\ref{fig:trivmod} should represent unbiased estimates of the
distribution of stellar mass among 2PIGG groups. 
The $\rf$-band curves are very similar to those
recovered from the mock catalogue, with the main differences being
that the real Universe contains a higher fraction of stars in very small
groups and groups with $\log_{10}(L_{b_{\rm j}})\sim10$. The model
stars are more frequently placed in groups with 
$\log_{10}(L_{b_{\rm j}})\sim9-9.5$ than is the case in the real Universe.
The difference for the smallest groups can be understood by recalling
that the mock catalogue 
did not include galaxies with $M_{b_{\rm J}}>-16$. With these
very low luminosity galaxies missing, fewer stars are placed
into the smallest groups. In the $J$-band, the results are broadly
similar to the $\rf$ band, with a slight deficit of stars
in small groups. This reflects the difference in faint-end
slopes of the $\rf$-inferred and $J$-inferred stellar mass functions
shown in Fig.~\ref{fig:schecmreal}, at least some of which results
from the fact that 2MASS misses some low surface brightness galaxies
(Andreon 2002; Bell \etal 2003). As this additional systematic bias is known
to affect the $J$-band results, the following fit is provided only for the
$\rf$-inferred stellar mass distribution:
\begin{equation}
f(M_{\rm stars})(>L_{b_{\rm J}})=1-\int_{-\infty}^{\log_{10}L_{b_{\rm
J}}}\frac{\exp[\frac{(x-\bar{x})^2}{2\sigma^2}]}{\sqrt{2\pi\sigma^2}}~\d x,
\end{equation}
where $\sigma=0.96$ and $\bar{x}=10.39$.
This fit to the fraction of stellar mass in haloes with a total blue
luminosity exceeding $L_{b_{\rm J}}$ is accurate to better than
$0.02$ for all group luminosities. 

\section{Conclusions}\label{sec:conc}

The near-IR light and stellar content of the Universe as a whole, and its
constituent groups, have been quantified using the 2dFGRS, the 
2PIGG catalogue and the 2MASS. This work extends that of
previous studies with the largest sets of galaxies and groups yet
used for these purposes. Furthermore, the mock galaxy catalogues that
have been employed, allow a careful quantification of the systematic errors
associated with estimating stellar masses. This represents an
important advance, because the systematic bias introduced by errors in
the estimated stellar masses, which has been quantified via the mock
catalogues, is sufficiently large that the mean stellar mass density
had previously been significantly overestimated. The mock catalogues
permit this bias to be quantified and corrected for.

The mean luminosity density in the Universe is  found to be
$\bar\rho_J=(3.57\pm0.11)\times10^8 h L_\odot{\rm Mpc}^{-3}$ and
$\bar\rho_{K_S}=(7.04\pm0.23)\times10^8 h L_\odot{\rm Mpc}^{-3}$
(statistical uncertainty only) in the two near-IR bands considered
here. Systematic uncertainties are likely to be of order $20$ per cent
and dominated by the completeness of the catalogues used and the
conversion of Kron to total magnitudes.
Taking into account the $\sim 40$ per cent overestimation of the
stellar mass density that, according to the mock catalogues, arises
from the uncertainty in inferring stellar 
masses from galaxy fluxes and colours, the mean stellar mass density
amounts to $\Omega_{\rm stars}h=(0.99\pm0.03)\times10^{-3}$. This value
assumes that a Kennicutt IMF is applicable to all sites of star formation.
If a Salpeter IMF had been chosen then the estimated $\Omega_{\rm stars}h$
would be a factor of $\sim 2.1$ larger.
The value of $\Omega_{\rm stars}h$ provides a measure of the density of
material currently locked up in stars. In conjunction with knowledge
about what fraction of stellar mass is recycled, this gives a
constraint on the integral over time of the Universal star formation rate.

The stellar mass-to-light ratio and stellar
mass fraction are both studied as functions of 2PIGG size. Rich
clusters are found to have stellar mass-to-$K_S$ band light ratios of
$\sim 0.6 \Upsilon_\odot$, a value that is about $60$ per cent larger
than that found to be typical of groups with 
$\log_{10}[L_{b_{\rm J}}/(\Lsol)]=10$. Removing the systematic errors
inherent in this determination reduces the cluster value to
$0.45\Upsilon_\odot$. The opposite trend is found for the stellar mass
fraction, with larger groups having smaller values. Taking into
account the systematic error in the recovery gives a typical stellar
mass fraction of $\sim5\times10^{-3}h$ for the richest clusters.

Finally, a trivariate stepwise maximum likelihood method is employed
to partition, into groups of different size, the stellar mass that resides
in galaxies in the local Universe. It is found that only a couple
of per cent of this stellar mass resides in galaxy clusters with
$\log_{10}[L_{b_{\rm J}}/(\Lsol)]>12$
($\log_{10}[M/(\Msol)]\gsim14.7$). Half of this stellar mass is
located in groups with $\log_{10}[L_{b_{\rm J}}/(\Lsol)]>10.4$
($\log_{10}[M/(\Msol)]\gsim12.5$). Once again, using mock
catalogues can reduce the systematic biases associated with
the entire measurement procedure. This adds weight to the assertion
that the measured distribution is an accurate representation of the
true underlying distribution of stellar mass among different sized
haloes and, as such, represents a valuable link in the chain
connecting theory with observation.

\section*{ACKNOWLEDGMENTS}

VRE and CMB are Royal Society University Research Fellows.
HMK was supported by a Royal Society Summer Studentship.
JAP is a PPARC Senior Research Fellow.
This work was supported by a PPARC rolling grant for Extragalactic
Astronomy and Cosmology. The 2dF Galaxy Redshift Survey was
made possible through the dedicated efforts of the staff of the
Anglo-Australian Observatory, both in creating the 2dF instrument and
in supporting it on the telescope. The 2dFGRS Team are also
acknowledged for their efforts in producing the survey.
This publication makes use of data products from the Two Micron All
Sky Survey, which is a joint project of the University of
Massachusetts and the Infrared Processing and Analysis
Center/California Institute of Technology, funded by the National
Aeronautics and Space Administration and the National Science
Foundation.

{}

\end{document}